\documentclass[a4paper,12pt]{article}
\usepackage{amsfonts}
\usepackage[xdvi]{graphicx}
\usepackage[dvips]{color}
\usepackage{amsmath,amssymb,amsthm}
\usepackage{textcomp}
\usepackage{ulem}

\makeatletter

\@addtoreset{equation}{section}

\newcounter{def}[section]
\renewcommand{\thedef}{\stepcounter{def}\thesection.\@arabic\c@def }

\begin{document}
\setlength{\baselineskip}{24pt}
\begin{center}
\textbf{\LARGE{Bifurcation of the neuronal population dynamics of the modified theta model: 
transition to macroscopic gamma oscillation }}
\end{center}

\setlength{\baselineskip}{14pt}

\begin{center}
Kiyoshi Kotani\footnote{kotani@neuron.t.u-tokyo.ac.jp}\\
Research Center for Advanced Science and Technology, The University of Tokyo, 4--6--1 Komaba, Meguro-ku, Tokyo 153-8904, Japan\\
\end{center}
\begin{center}
Akihiko Akao\footnote{aka64@pitt.edu}\\
Department of Mathematics, University of Pittsburgh, Pittsburgh, Pennsylvania 15213, USA \\
and Graduate School of Engineering, The University of Tokyo, 7--3--1 Hongo, Bunkyo-ku, Tokyo 113-0033, Japan\\
\end{center}
\begin{center}
Hayato Chiba\footnote{hchiba@tohoku.ac.jp}\\
Advanced Institute for Materials Research, Tohoku University, \\ 2-1-1 Katahira, Aoba-ku, Sendai, 980-8577, Japan 
\end{center}

 \begin{center}
 \end{center}

\begin{center}
\textbf{Abstract}
\end{center}
Interactions of inhibitory neurons produce gamma oscillations (30--80 Hz) in the local field potential, 
which is known to be involved in functions such as cognition and attention. 
In this study, the modified theta model is considered to investigate the theoretical relationship 
between the microscopic structure of inhibitory neurons and their gamma oscillations
under a wide class of distribution functions of tonic currents on individual neurons. 
The stability and bifurcation of gamma oscillations for the Vlasov equation of the model
is investigated by the generalized spectral theory.
It is shown that as a connection probability of neurons increases, 
a pair of generalized eigenvalues crosses the imaginary axis twice,
which implies that a stable gamma oscillation exists only when the connection probability has a value within a suitable range.
On the other hand, when the distribution of tonic currents on individual neurons is
the Lorentzian distribution, the Vlasov equation is reduced to a finite dimensional dynamical system. 
The bifurcation analyses of the reduced equation exhibit equivalent results with the generalized spectral theory. 
It is also demonstrated that the numerical computations of neuronal population follow the analyses of the generalized spectral theory as well as the bifurcation analysis of the reduced equation.


\section{Introduction}
The local field potential known as collective macroscopic oscillations in the brain
is organized by interacting neurons.
Among them, gamma oscillations (30--80 Hz) in the local field potential are known to be involved in functions such as cognition and attention\cite{Wang0, Womelsdorf, Rod, Singer}. 
In a physiological point of view, inhibitory synaptic interactions introduced by GABAergic neurons play important role to generate gamma oscillations\cite{Wang0, Bartos}.  
However, there are only limited approaches on how the properties of neurons at the microscopic level affect macroscopic gamma oscillations.

In this study, the modified theta model (MT model)~\cite{Kotani, Kotani2}, which is a dynamical system 
of membrane potentials of interacting neurons, is considered to investigate the theoretical 
relationship between the microscopic structure of inhibitory neurons and their gamma oscillations.
The Vlasov equation of the model is introdueced under the assumption that 
the network structure of neurons is the Erd\"{o}s-R\'{e}nyi random graph with the probability of connection $p$.
For the Vlasov equation, bifurcations from the steady state (de-synchronized state)
to the collective gamma oscillations (synchronized state) will be studied as the parameter $p$ varies under a wide class of distribution functions of tonic currents on individual neurons.

The stability and bifurcation analysis is not straightforward because of the continuous spectrum.
Let $T$ be a linear operator obtained by the linearization of the Vlasov equation around the steady state.
Since the operator $T$ has the continuous spectrum on the imaginary axis,
the standard stability theory of dynamical systems is not applicable.
To handle such a difficulty caused by the continuous spectrum, the generalized spectral theory 
will be employed which was developed to treat a similar problem for the Kuramoto model~\cite{Chi2, Chi3}.
Although there are no eigenvalues of $T$ when $p$ is small enough, it will be shown that there exist 
generalized eigenvalues on the left half plane.
We further assume that the variance of tonic currents on individual neurons is sufficiently small. 
Then, as the parameter $p$ increases, the generalized eigenvalues cross the continuous spectrum on 
the imaginary axis and it implies that the gamma oscillation occurs.
Further as $p$ increases, the eigenvalues again cross the continuous spectrum from 
the right half plane to the left half plane.
As a result, the gamma oscillation is destroyed.
These results imply that if the edge density is too large, the gamma oscillation is not realized.

In Sec.2, the modified theta model and its Vlasov equation are introduced.
A few properties of the steady state will be shown.
In Sec.3, the Vlasov equation is reduced to a certain system of evolution equations by Ott-Antonsen reduction.
The linearized system around the steady state with the linear operator $T$ will be derived.
In Sec.4, the distribution function for tonic currents on individual neurons is assumed to be 
the delta function.
In this case, the eigenvalue problem of the operator $T$ is completely resolved.
If the distribution function is not the delta function, the operator $T$ has the continuous spectrum in general,
which is studied in Sec.5.
In Sec.6, the generalized spectral theory is applied to investigate the stability and bifurcations of the steady state.
In Sec.7  we consider the Lorentzian distribution of tonic currents where the Vlasov equation is reduced to a finite dimensional dynamical system. 
It is demonstrated that the numerical computations of neuronal population follow the analyses of the generalized spectral theory as well as the bifurcation analysis of the reduced equation. 

\section{The modified theta model and its steady state}

For the dynamics of the membrane potential $V(t)$ of a Type 1 single neuron, 
the following type of quadratic integrate and fire (QIF) model with resting potential and threshold~\cite{Gerstner} is sometimes used
\begin{eqnarray} 
c_m\frac{dV}{dt} = g_L \frac{(V - V_R)(V - V_T)}{V_T-V_R} + I,
\end{eqnarray}
where $c_m,V_R, V_T, g_L$ and $I$ are membrane capacitance, resting potential, threshold potential, leak conductance and input current, respectively.
Values of these numerical constants used for numerical simulations in this paper are listed in Table 1,
although for mathematical statement, these specific values are not used.
When $I=0$, the system has two fixed points $V=V_R, V_T$, and most orbits approach to the stable one $V_R$
(if $V_R < V_T$).
On the other hand, if $I$ is larger than some value, two fixed points disappear by a saddle-node bifurcation and 
any solution diverges to $+\infty$ in a finite time.
We assume that a neuron fires when $V(t)$ reaches $+\infty$.
Then, $V(t)$ is reset to $-\infty$ and again goes to $+\infty$ periodically.

Next, we consider the large population of neurons with interactions
\begin{eqnarray} 
c_m \frac{dV_i}{dt} = g_L \frac{(V_i - V_R)(V_i - V_T)}{V_T-V_R} + I_i - g^i_{syn}(t) (V_i-V_{syn}),
\quad i=1,\cdots ,N,
\label{2-2}
\end{eqnarray}
where $V_{syn}$ is a reversal potential and $g^i_{syn}(t)$ is an $i$-th synaptic conductance governed by the equation
\begin{eqnarray} 
\frac{dg^i_{syn}}{dt} = -\frac{1}{\tau } g^i_{syn} + g_{peak}\sum_{conn} \sum_{fire} \delta (t-t_{k}^j).
\label{2-3}
\end{eqnarray}
Here, $g_{peak}$ and $\tau$ denote a peak conductance and a decay time, respectively (see also Table 1).
The summation $\sum_{conn}$ runs over all neurons connected with the $i$-th neuron.
Thus, it reflects the structure of a network of neurons.
The second summation $\sum_{fire}$ runs over all times $t^j_k$ at which a $j$'s connected neuron gets firing.
Hence, a neuron affects the dynamics of neighboring neurons when it fires.
Eq.(\ref{2-2}) is known as the Riccati equation which is extended to a smooth vector field on a compactified
phase space $S^1$ (circle), on which $V_i=+\infty$ is topologically connected with $-\infty$.
Specifically, we introduce the coordinate transformation $V_i \mapsto \theta _i$ of the form
\begin{equation}
V_i = \frac{V_R + V_T}{2} + \frac{V_T - V_R}{2}\tan \frac{\theta _i}{2}.
\end{equation}
Then, Eq.(\ref{2-2}) is rewritten as the MT model introduced in Kotani et al. \cite{Kotani, Aki}
\begin{eqnarray} 
c_m \frac{d\theta _i}{dt} = -g_L \cos \theta _i + c_1(1+\cos \theta _i) I_i
 + g^i_{syn} (t) (c_2( 1 + \cos \theta _i) - \sin \theta _i) ,
\label{2-5}
\end{eqnarray}
where $c_1 = 2/(V_T-V_R)$ and $c_2 = (2V_{syn} - V_R - V_T)/(V_T - V_R)$ are constants.
By this transformation, $V = V_T, V_R, (V_T + V_R)/2$ and $\pm \infty$ are mapped to points
$\theta =\pi/2, -\pi/2, 0$ and $\pi$ on a circle, respectively. 

Without loss of generality, we can drop $c_m$ by rescaling time 
(Indeed, typical physiological value of $c_m$ is known to be about 1~\textmu F/cm$^2$). 
For mathematical analysis of the dynamics, 
it is convenient to consider the continuous limit (Vlasov equation) of (\ref{2-5}) given by
\begin{eqnarray} 
\left\{ \begin{array}{ll}
\displaystyle \frac{\partial P}{\partial t} + \frac{\partial }{\partial \theta }(vP) = 0,\,\,P=P(t, \theta , I) &  \\[0.4cm]
v(t, \theta , I) =  -g_L \cos  \theta  + c_1(1+\cos \theta) I
 + g_{syn} (t) (c_2( 1 + \cos \theta ) - \sin \theta ),  &  \\
\end{array} \right.
\end{eqnarray}
where $P(t,\theta ,I)$ is a probability density of neurons on $S^1$ parameterized by the current $I$ and time $t$.
Further, we assume that a network is the Erd\"{o}s-R\'{e}nyi random graph with the probability of connection $p$,
and consider the homogenization (averaging with respect to the realization of random graphs) 
of Eq.(\ref{2-3}) given by 
\begin{eqnarray} 
\frac{dg_{syn}}{dt} = -\frac{1}{\tau } g_{syn} + g_{peak}\cdot p \cdot N  \int_{\mathbf{R}}\! v(t, \pi, I) P(t, \pi, I) G(I) dI. 
\label{2-7}
\end{eqnarray}
$N$ denotes the number of neurons, thus the first summation $\sum_{conn}$ is replaced by the number of edges $pN$.
The second summation $\sum_{fire}$ is replaced by the above integral 
that denotes the flux of neurons at $\theta =\pi$ at which they fire. 
Here, we suppose that the current $I$ is a random variable drawn from a given distribution function $G(I)$.
We introduce the following notation
\begin{eqnarray} 
\left\{ \begin{array}{l}
\displaystyle f(I) = -\frac{1}{2}g_L + \frac{1}{2}c_1I + \frac{1}{2}c_2 g_{syn}(t) + \frac{i}{2}g_{syn}(t) \\[0.3cm]
\displaystyle h(I) = c_1I + c_2 g_{syn}(t) \\[0.2cm]
\displaystyle A(t) = \int_{\mathbf{R}}\! v(t, \pi, I) P(t, \pi, I) G(I) dI = g_L  \int_{\mathbf{R}}\! P(t, \pi, I) G(I)dI \\[0.2cm]
\mu = g_{peak}\cdot p \cdot N, 
\end{array} \right.
\label{2-8}
\end{eqnarray}
where $A(t)$ is a firing rate of the population and the relation $v(t,\pi, I)=g_L$ is used.
Then, our system is written as
\begin{eqnarray} 
\left\{ \begin{array}{ll}
\displaystyle \frac{\partial P}{\partial t} + \frac{\partial }{\partial \theta }(vP) = 0,\,\,P=P(t, \theta , I) &  \\[0.4cm]
v(t, \theta , I) =  f(I) e^{i\theta } + h(I) + \overline{f(I)}e^{-i\theta } &  \\[0.3cm]
\displaystyle \frac{dg_{syn}}{dt} = -\frac{1}{\tau } g_{syn} + \mu A(t).  &
\end{array} \right.
\label{2-9}
\end{eqnarray}
In particular, $\mu =  g_{peak} \cdot p \cdot N$ controls the edge density of the network.
Our purpose is to investigate bifurcations of the system from the steady state (de-synchronous state) to
the synchronous collective firing (gamma oscillation) as the bifurcation parameter $\mu$ varies.

\begin{table}[h]
\begin{center}
\begin{tabular}{|c|c|l|} 
\hline
$V_i(t)$ & membrane potential & unknown variable \\ \hline
$g_{syn}^i (t)$ & synaptic conductance & unknown variable \\ \hline
$I_i$ & input current & i.i.d. drawn from a distribution $G(I)$ \\ \hline 
$c_m$ & membrane capacitance & 1 (\textmu F/cm$^2$) \\ \hline 
$V_R$ & resting potential & $-62$ (mV) \\ \hline
$V_T$ & threshold potential & $-55$ (mV) \\ \hline
$V_{syn}$ & reversal potential & $-70$ (mV) \\ \hline
$g_L$ & leak conductance & 0.1 (mS/cm$^2$) \\ \hline
$g_{peak}$ & peak conductance & 0.0214 (mS/cm$^2$) \\ \hline
$\tau$ & decay time & 5 (ms) \\ \hline
$c_1$ & $2/(V_T-V_R)$ & 2/7 \\ \hline
$c_2$ & $ (2V_{syn}\! -\! V_R \!-\! V_T)/(V_T - V_R)$ & $-23/7$ \\ \hline
$N$ & the number of neurons & 800 \\ \hline
\end{tabular}
\caption{Names of variables and values used for numerical simulations.}
\end{center}
\end{table}

In what follows, we assume that
\\
(1) $g_L, g_{peak}, c_1$ are positive and $c_2$ is a negative number
(for numerical simulations, we use the specific values shown in Table 1. 
They are physiologically plausible for GABAergic neurons~\cite{Kotani}), and
\\
(2) $G(I)$ is an even and unimodal function around the average $\eta$, which implies
$G(I+\eta) = G(-I + \eta)$ and $G'(I) < 0\, (I>\eta)$.
\\

Let us derive the steady state (de-synchronous state) $(P, g_{syn}) = (P_0, g_0)$ of the system (\ref{2-9}).
We need the following lemma.
\\

\textbf{Lemma \thedef.} For the quadratic equation
\begin{eqnarray*}
fz^2 + hz + \overline{f} = 0, \quad f\in \mathbf{C},\, h\in \mathbf{R},
\end{eqnarray*}
the roots
\begin{eqnarray*}
z_{\pm} = \frac{1}{2f}\left( -h \pm \sqrt{h^2 - 4|f|^2} \right)
\end{eqnarray*}
satisfy the following properties.
\\[0.2cm]
(i) When $h^2 - 4|f|^2 \leq 0$, $|z_{\pm}| = 1$. \\
(ii) When $h^2 - 4|f|^2 > 0$, $|z_+| < 1$ and $|z_-| > 1$.
\\

Let $(f_0(I), h_0(I))$ denotes $(f(I), h(I))$ for which the value of the steady state $g_{syn} (t) = g_0$ is substituted.
Define two sets
\begin{eqnarray*}
\Omega _1 &=& \{ I \in \mathbf{R} \, | \, h_0(I)^2 - 4|f_0(I)|^2 > 0\} \\
 \Omega _2 &=& \{ I \in \mathbf{R} \, | \, h_0(I)^2 - 4|f_0(I)|^2 \leq 0\}.
\end{eqnarray*}
Due to Lemma 2.1, the equation $f_0(I) e^{i\theta } + h_0(I) + \overline{f_0(I)}e^{-i\theta }=0$ has real solution only when $I\in \Omega _2$.
Hence, using unknown $g_0$, the steady state of the first equation of (\ref{2-9}) is given by
\begin{equation}
P_0(\theta , I) = \left\{ \begin{array}{ll}
\displaystyle \frac{C(I)}{f_0(I) e^{i\theta } + h_0(I) + \overline{f_0(I)}e^{-i\theta }} & (I\in \Omega _1) \\[0.5cm]
\displaystyle \delta (\theta -\theta _0(I)) & (I\in \Omega _2), \\
\end{array} \right.
\label{2-10}
\end{equation}
where $C(I)$ is a normalization constant and 
 $\delta (\theta -\theta _0(I))$ is the delta function supported at $\theta _0(I)$.
 $\theta _0(I)$ satisfies $f_0(I) e^{i\theta _0(I) } + h_0(I) + \overline{f_0(I)}e^{-i\theta _0(I) }=0$ and $-\pi<\theta _0(I)\leq 0$.  
The normalization condition provides
\begin{eqnarray*}
1 = \int^{2\pi}_{0}\! \frac{C(I)}{f_0(I) e^{i\theta } + h_0(I) + \overline{f_0(I)}e^{-i\theta }} d\theta .
\end{eqnarray*}
Changing variables $e^{i\theta } = z$ yields
\begin{eqnarray*}
1 = \oint\! \frac{-iC(I)}{f_0(I) z^2 + h_0(I)z + \overline{f_0(I)}} dz
 = \oint\! \frac{-iC(I)}{f_0(I) (z - z_+)(z-z_-)}dz,
\end{eqnarray*}
where the integral path is the unit circle.
Lemma 2.1 shows $z_+$ is inside the unit circle while $z_-$ is not.
Thus, the residue theorem gives
\begin{eqnarray*}
1 = \frac{2\pi C(I)}{f_0(I)}\frac{1}{z_+ - z_-}\quad \Rightarrow \quad C(I) = \frac{1}{2\pi} \sqrt{h_0(I)^2 - 4|f_0(I)|^2}.
\end{eqnarray*}
The next task is to calculate $g_0$.
When $P$ is in the steady state $P_0(\theta ,I)$, we have
\begin{eqnarray*}
A(t) = g_L \int_{\mathbf{R}}\! P_0(\pi, I) G(I)dI
 = \int_{\Omega _1}\! C(I)G(I)dI + g_L \int_{\Omega _2}\! \delta (\pi - \theta _0(I)) G(I)dI.  
\end{eqnarray*}
Since $\theta _0(I) \neq \pi$, the second term in the right hand side disappears.
The steady state $g_{syn}(t) = g_0$ is given by a solution of the following equation
\begin{eqnarray}
g_0 =  \tau \mu A(t) = \frac{\tau \mu}{2\pi} \int_{\Omega _1}\! \sqrt{h_0(I)^2 - 4|f_0(I)|^2} G(I)dI.
\label{2-11}
\end{eqnarray}
Changing the variables from $I$ to $x$ by the relation
\begin{equation}
h_0(I)^2 - 4|f_0(I)|^2 = 2g_L(c_1 I + c_2 g_0) - g_0^2 - g_L^2 =: x
\label{2-12}
\end{equation} 
yields
\begin{equation}
g_0 = \frac{\tau \mu}{4\pi g_L c_1} \int^{\infty}_{0}\! \sqrt{x}\cdot G \left( \frac{x + g_0^2 + g_L^2 - 2c_2g_Lg_0}{2g_Lc_1}\right) dx.
\label{2-13}
\end{equation}
Further change of variables by $x = \omega ^2$ gives
\begin{equation}
g_0 = \frac{\tau \mu}{4\pi g_L c_1} \int^{\infty}_{-\infty}\! \omega ^2 \cdot \widetilde{G} \left( \omega \right) d\omega,
\label{2-14}
\end{equation}
where $\widetilde{G}$ is an even function defined by
\begin{equation}
\widetilde{G}(\omega ) = G\left( \frac{\omega ^2 + g_0^2 + g_L^2 - 2c_2g_Lg_0}{2g_Lc_1}\right).
\label{2-15}
\end{equation}
The steady state (de-synchronous state) $(P_0, g_0)$ is given by (\ref{2-10}) and a solution of (\ref{2-13}) or (\ref{2-14}),
though $g_0$ can be calculated only numerically.
\\

\textbf{Proposition \thedef.} Suppose that $G(I)$ is even and unimodal around the average $\eta$.
For any $\mu \geq 0$, a nonnegative solution $g_0 = g_0(\mu)$ of (\ref{2-13}) uniquely exists.
$g_0(\mu)$ is monotonically increasing in $\mu$ and $g_0(\mu) \to \infty$ as $\mu \to \infty$.
\\

\textbf{Proof.} Put
\begin{eqnarray*}
F(z) = \frac{\tau}{4\pi g_L c_1}\int^{\infty}_{0}\! \sqrt{x}\cdot G \left( \frac{x + z^2 - 2c_2g_Lz + g_L^2 }{2g_Lc_1}\right) dx.
\end{eqnarray*}
Then, $g_0$ is given by the intersection of the line $y = z /\mu$ and the curve $y = F(z)$.
Since $F(z) > 0$ for any $z \geq 0$, it is sufficient to show that $F'(z) < 0$ for $z>0$.
Then, $F(z)$ is monotonically decreasing and Proposition is proved.
\begin{eqnarray*}
F'(z) = \frac{\tau}{4\pi g_L c_1} \frac{2z-2c_2g_L}{2c_1g_L} \int^{\infty}_{0}\! 
\sqrt{x}\cdot G' \left( \frac{x + z^2 - 2c_2g_Lz + g_L^2 }{2g_Lc_1}\right) dx.
\end{eqnarray*}
Because of our assumption for the signs of parameters, the number in the front of the integral is positive
(recall that $c_2 < 0$).

By the assumption for $G$, $G'(I)$ is an odd function around $I=\eta$ satisfying 
$G'(I) < 0$ for $I>\eta$ and $G'(\eta ) = 0$.
Thus, there is a unique number $x_0$ such that 
\begin{eqnarray*}
H(x) = G' \left( \frac{x + z^2 - 2c_2g_Lz + g_L^2 }{2g_Lc_1}\right)
\end{eqnarray*}
is an odd function around $x = x_0$ satisfying $H(x) < 0$ for $x > x_0$ and $H(x_0) = 0$.
If $x_0 \leq 0$, then $H(x) < 0$ for $x>0$ and $F'(z)$ is negative.
If $x_0 > 0$, we decompose the above integral as
\begin{eqnarray*}
\int^{\infty}_{0}\! \sqrt{x} H(x)dx = \int^{x_0}_{0}\! \sqrt{x} H(x)dx
+\int^{2x_0}_{x_0}\! \sqrt{x} H(x)dx + \int^{\infty}_{2x_0}\! \sqrt{x} H(x)dx.
\end{eqnarray*}
By the assumption, the sum of the first two terms is negative, and the third term is also negative,
which proves that $F'(z) < 0$.


\section{Ott-Antonsen reduction}
For the system (\ref{2-9}), we expand a solution in a Fourier series.
Putting 
\begin{eqnarray*}
z_k(t, I) = \int^{2\pi}_{0}\! P(t,\theta ,I) e^{ik\theta }d\theta,
\end{eqnarray*}
we rewrite (\ref{2-9}) as the system of equations of $z_k(t,I)$'s given by 
\begin{eqnarray}
\left\{ \begin{array}{l}
\dot{z}_1= i(\, f(I) z_2 + h(I) z_1 + \overline{f(I)}\,)  \\[0.2cm]
\dot{z}_k= ik(\, f(I) z_{k+1} + h(I) z_k + \overline{f(I)} z_{k-1}\,)  \\[0.2cm]
\displaystyle \dot{g}_{syn} = -\frac{1}{\tau} g_{syn} + \mu A(t) \\[0.2cm]
\displaystyle A(t) = \frac{g_L}{2\pi} \int_{\mathbf{R}}\! \left( 1 + \sum^\infty_{k=1} (z_{2k} + \overline{z}_{2k})
 - \sum^\infty_{k=1} (z_{2k-1} + \overline{z}_{2k-1}) \right) G(I)dI. 
\end{array} \right.
\end{eqnarray}
It is easy to verify that the set in the phase space defined by $\{ z_k = z_1^k\, | \, k=1,2\cdots \}$
is an invariant set of this system, on which it is reduced to the system (Ott-Antonsen reduction~\cite{OA,Paul1, Paul2}). Thus, the reduced equations, which has been introduced in Ref.\cite{Aki}, are given by
\begin{eqnarray} 
\left\{ \begin{array}{l}
\dot{z}_1= i(\, f(I) z_1^2 + h(I) z_1 + \overline{f(I)}\,)  \\[0.2cm]
\dot{\overline{z}}_1 = -i (\,\overline{f(I)}\,\, \overline{z}_1^2 + h(I) \overline{z}_1 + f(I)\,) \\[0.2cm] 
\displaystyle \dot{g}_{syn} = -\frac{1}{\tau}g_{syn} + \mu A(t) \\[0.2cm]
\displaystyle A(t) = \frac{g_L}{2\pi} \int_{\mathbf{R}} \left( 1 - \frac{z_1}{1+z_1} - \frac{\overline{z}_1}{1+\overline{z}_1}\right) G(I)dI. 
\end{array} \right.
\label{3-2}
\end{eqnarray}
We regard this system as a dynamical system on the Hilbert space 
$\mathcal{H}:=L^2(\mathbf{R}, G(I)dI) \oplus L^2(\mathbf{R}, G(I)dI)\oplus \mathbf{R} $
(the equation for the complex conjugate $\overline{z}_1$ is added 
because $L^2(\mathbf{R}, G(I)dI)$ is considered as an $\mathbf{R}$-vector space).
\\

\textbf{Lemma \thedef.} The steady state $(P_0, g_0)$ shown in Section 2 lies on the invariant set.
\\

\textbf{Proof.} We show that the steady state satisfies $z_k = z_1 ^k$. When $I\in \Omega _2$, 
\begin{eqnarray*}
z_k = \int^{2\pi}_{0}\! \delta (\theta -\theta _0(I)) e^{ik\theta } d\theta = e^{ik\theta _0(I)} = z_1^k.
\end{eqnarray*}
When $I\in \Omega _1$, 
\begin{eqnarray*}
z_k = \int^{2\pi}_{0}\! \frac{C(I)}{f_0(I)e^{i\theta } + h_0(I) + \overline{f_0(I)}e^{-i\theta }} e^{ik\theta }d\theta. 
\end{eqnarray*}
As in the previous section, this integral is calculated by using the residue theorem with the definition of $C(I)$ as
\begin{eqnarray*}
z_k = \frac{2\pi C(I)}{f_0(I)}\frac{z_+^k}{z_+ -z_-} = z_+^k = z_1^k.
\end{eqnarray*}
This proves the lemma.
\\

In what follows, we investigate the linear stability of the steady state $(z_1, \overline{z}_1, g_0) = 
(z_+(I), \overline{z_+(I)}, g_0)$ of Eq.(\ref{3-2}) given by
\begin{eqnarray*}
 z_+(I) = \frac{1}{2f_0(I)} \left( -h_0(I) + \sqrt{h_0(I)^2 - 4|f_0(I)|^2} \right)
\end{eqnarray*}
and a positive solution of (\ref{2-13}).
The linearized equation of (\ref{3-2}) around this steady state is
\begin{eqnarray} 
\frac{d}{dt} \left(
\begin{array}{@{\,}c@{\,}}
u \\
\overline{u} \\
v
\end{array}
\right) = \left(
\begin{array}{@{\,}ccc@{\,}}
Q(I) & 0 & P(I) \\
0 & \overline{Q(I)} & \overline{P(I)} \\
R(I) & \overline{R(I)} & -1/\tau 
\end{array}
\right) 
\left(
\begin{array}{@{\,}c@{\,}}
u \\
\overline{u} \\
v
\end{array}
\right)
:= T \left(
\begin{array}{@{\,}c@{\,}}
u \\
\overline{u} \\
v
\end{array}
\right),
\label{3-3}
\end{eqnarray}
where
\begin{eqnarray*}
\left\{ \begin{array}{l}
Q(I) = i \sqrt{h_0(I)^2 - 4|f_0(I)|^2}, \\
\displaystyle P(I) = \frac{i}{2}(1+z_+)((c_2 + i)z_+ + c_2 - i), \\
\displaystyle R(I)u = -\frac{\mu g_L}{2\pi}\int_{\mathbf{R}}\! \frac{u}{(1+z_+)^2}G(I)dI.
\end{array} \right.
\end{eqnarray*}
$Q(I)$ and $P(I)$ are multiplication operators on $L^2(\mathbf{R}, G(I)dI)$ and $\mathbf{R}$, respectively,
and $R(I)$ is the integral operator on $L^2(\mathbf{R}, G(I)dI)$.
Thus, the above matrix defines a linear operator $T$ on $\mathcal{H}$.
For the linear stability of the steady state, we will investigate the spectrum of $T$. 
The following relations will be often used.
\begin{eqnarray}
& & h_0(I)^2 - 4|f_0(I)|^2 = 2g_L(c_1 I + c_2 g_0) - g_0^2 - g_L^2, \label{3-4} \\
& & \frac{P(I)}{(1+z_+)^2} = \frac{1}{2g_L} \left( \sqrt{h_0(I)^2 - 4|f_0(I)|^2} + i (c_2g_L - g_0) \right) .\label{3-5}
\end{eqnarray}


\section{Delta distribution}

First, we consider the simplest case: we suppose that the distribution $G(I) = \delta (I-\eta)$ is the 
delta function supported at $\eta$.
In this case, the integral operator is not involved and the system (\ref{3-2}) is 
reduced to a 3-dim system.
The equation (\ref{2-11}) to obtain $g_0$ is reduced to 
\begin{eqnarray*}
g_0 = \frac{\mu \tau}{2\pi} \sqrt{h(\eta)^2 - 4|f(\eta)|^2} = \frac{\mu \tau}{2\pi} \sqrt{2g_L(c_1 \eta + c_2 g_0) - g_0^2 - g_L^2}.
\end{eqnarray*}
The eigen-equation of the linearized system around the steady state $(z_+(\eta), \overline{z_+(\eta)}, g_0)$ is given by
\begin{eqnarray}
f(\lambda ):= \mathrm{det} \left(
\begin{array}{@{\,}ccc@{\,}}
\lambda - Q(\eta) & 0 & -P(\eta) \\
0 & \lambda -\overline{Q(\eta )} & -\overline{P(\eta)} \\
\displaystyle \frac{\mu g_L}{2\pi}\frac{1}{(1+z_+(\eta ))^2} 
  & \displaystyle \frac{\mu g_L}{2\pi}\frac{1}{(1+\overline{z_+(\eta)})^2} & \lambda +1/\tau
\end{array}
\right) =0.
\label{4-1}
\end{eqnarray}
When $\mu = 0$, $g_0 = 0$. Thus, three eigenvalues are given by $\lambda  = -1/\tau$ and $\lambda  = \pm i \sqrt{2g_Lc_1\eta - g_L^2}$.
Since we are interested in the macroscopic oscillation which will appear through a Hopf bifurcation,
we assume that $2c_1 \eta - g_L > 0$; i.e. there exist two eigenvalues on the imaginary axis when $\mu = 0$.
\\

\textbf{Proposition \thedef.} Suppose $\eta > g_L/(2c_1)$ and $1/\tau < - c_2 g_L$.
For any $\mu > 0$, $f(\lambda ) = 0$ has a negative real root $\alpha $ and a pair of complex roots
$\beta, \gamma = \overline{\beta}$ whose real parts are positive.
The pair transversely crosses the imaginary axis at $\mu = 0$ from the left half plane to the right half plane.
\\

\textbf{Proof.} As in Eq.(\ref{2-12}), we put 
\begin{eqnarray*}
x = 2g_L(c_1 \eta + c_2 g_0) - g_0^2 - g_L^2 = \left( \frac{2\pi g_0}{\tau \mu} \right)^2.
\end{eqnarray*}
With the aid of Eqs.(\ref{3-4}) and (\ref{3-5}), we can show that the eigen-equation (\ref{4-1}) is rearranged as 
\begin{eqnarray*}
f(\lambda ) = \lambda ^3 + \frac{1}{\tau}\lambda ^2 + (x+\frac{1}{\tau} g_0) \lambda + \frac{1}{\tau} (x + g_0^2 - c_2 g_Lg_0) = 0
\end{eqnarray*}
The relation between roots and coefficients implies
\begin{eqnarray*}
\alpha + \beta + \gamma = -\frac{1}{\tau}, \quad
\alpha \beta +\beta \gamma + \gamma \alpha = x + \frac{1}{\tau}g_0, \quad
\alpha \beta \gamma  = -\frac{1}{\tau} (x + g_0^2 - c_2 g_Lg_0).
\end{eqnarray*}
Since all coefficients are positive, it turns out that there are no positive real roots.
By the assumption, we have
\begin{eqnarray*}
f(-1/\tau) = \frac{g_0}{\tau} (g_0 - \frac{1}{\tau} - c_2 g_L) > 0,
\end{eqnarray*}
which implies that there exists a negative real root $\alpha $ satisfying $\alpha < -1/\tau$.
This shows $\beta + \gamma = -1/\tau - \alpha >0$.
Hence, if $\beta$ and $\gamma $ are real, at least one of them is positive.
This is a contradiction and they should be a complex pair with positive real parts.
The last statement follows from the implicit function theorem to $f(\lambda ) = 0$.
\\

\begin{figure}
\begin{center}
\includegraphics[scale=0.5]{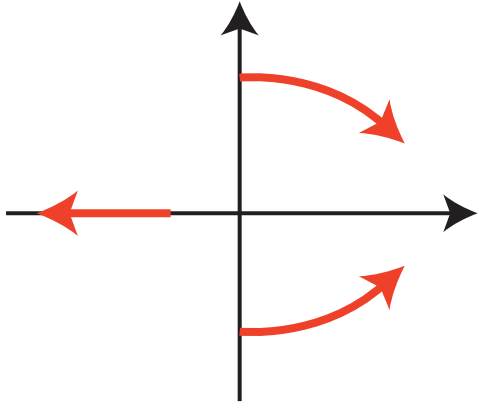}
\caption[]{(Color online)The motions of eigenvalues as $\mu$ increases from zero.}
\label{fig1}
\end{center}
\end{figure}

Fig.\ref{fig1} represents the motions of eigenvalues as $\mu$ increases from zero.
The proposition implies that a Hopf bifurcation occurs at $\mu = 0$ and 
the gamma oscillation is stable for any $\mu > 0$.

\section{Stability analysis 1}

\begin{figure}
\begin{center}
\includegraphics[scale=0.5]{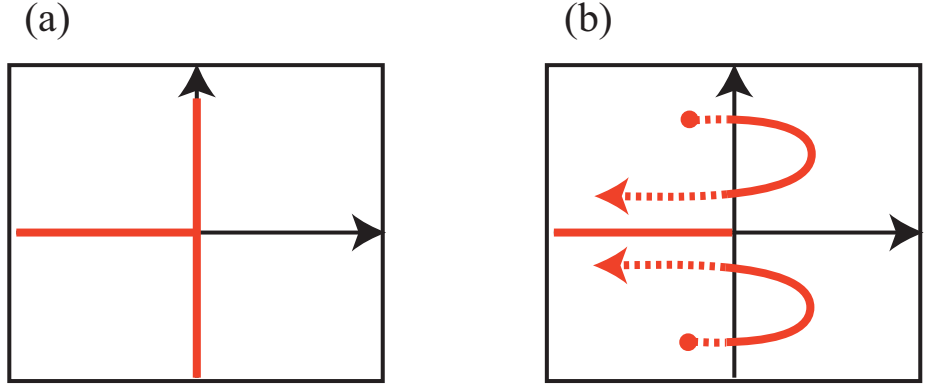}
\caption[]{(Color online)(a) The continuous spectrum of the operator $T$. (b) The generalized spectrum of $T$.
The dotted lines denote the motion of generalized eigenvalues and solid lines denote that of usual eigenvalues
as $\mu$ increases.
See Section 6.}
\label{fig2}
\end{center}
\end{figure}

In general, the system (\ref{3-2}) is an infinite dimensional dynamical system on $\mathcal{H}$
and the linear operator $T$ has the continuous spectrum.
Let us derive the eigen-equation of the operator $T$ defined by (see Eq.(\ref{3-3}))
\begin{eqnarray*}
 \left(
\begin{array}{@{\,}ccc@{\,}}
\lambda -Q(I) & 0 & -P(I) \\
0 & \lambda -\overline{Q(I)} & -\overline{P(I)} \\
-R(I) & -\overline{R(I)} & \lambda +1/\tau 
\end{array}
\right) 
\left(
\begin{array}{@{\,}c@{\,}}
u \\
\overline{u} \\
v
\end{array}
\right) = 0.
\end{eqnarray*}
The first two equations give $u = (\lambda - Q(I))^{-1} P(I) v$.
By substituting it into the third equation, we obtain
\begin{eqnarray}
\lambda + \frac{1}{\tau} = -\frac{\mu g_L}{2\pi}\! \int_{\mathbf{R}}\! \left(  
\frac{P(I)}{(1+z_+)^2}\frac{1}{\lambda -Q(I)} + \frac{\overline{P(I)}}{(1+\overline{z_+})^2 }\frac{1}{\lambda -\overline{Q(I)}} 
\right) \! G(I)dI .
\label{5-1}
\end{eqnarray}
The factors $(\lambda -Q(I))^{-1}$ and $(\lambda - \overline{Q(I)})^{-1}$ suggest that
if the support of $G$ is the whole real axis, which will be assumed in Sec.6, the continuous spectrum of $T$ is given by the set
\begin{equation}
\{ Q(I) \, | \, I\in \mathbf{R}\} \cup \{ \overline{Q(I)} \, | \, I\in \mathbf{R}\} =i \mathbf{R} \cup \mathbf{R}_{<0},
\end{equation}
that is, the whole imaginary axis and the negative real axis (Fig.\ref{fig2}).
Because of the spectrum on the imaginary axis, the steady state looks neutrally stable in $\mathcal{H}$-topology.
In the next section, we will employ the generalized spectral theory to treat the continuous spectrum. 

We further simplify Eq.(\ref{5-1}).
When $I\in \Omega _2$, $Q(I) = \overline{Q(I)}$ is a real number.
We also have
\begin{eqnarray*}
& & \frac{P(I)}{(1+z_+)^2} + \frac{\overline{P(I)}}{(1+\overline{z_+})^2} \\
&= & \frac{i}{2} \left( \frac{(c_2 + i)z_+ + c_2 - i}{1+z_+}-\frac{(c_2 - i)\overline{z_+} + c_2 + i}{1+\overline{z_+}} \right)
 = \frac{1-|z_+|^2}{|1+z_+|^2}.
\end{eqnarray*}
Due to Lemma 2.1, $|z_+| = 1$ and the above quantity becomes zero.
Thus, we can assume $I\in \Omega _1$ in Eq.(\ref{5-1}): 
\begin{eqnarray*}
\lambda + \frac{1}{\tau} = -\frac{\mu g_L}{2\pi}\! \int_{\Omega _1}\! \left(  
\frac{P(I)}{(1+z_+)^2}\frac{1}{\lambda -Q(I)} + \frac{\overline{P(I)}}{(1+\overline{z_+})^2 }\frac{1}{\lambda -\overline{Q(I)}} 
\right) \! G(I)dI .
\end{eqnarray*}
Changing the variables by (\ref{2-12}) and using the formula (\ref{3-5}) yield
\begin{eqnarray*}
&&\lambda + \frac{1}{\tau} =\\
&& -\frac{\mu }{4\pi c_1}\! \int^\infty_{0}\! \frac{1}{2g_L}\left( 
\frac{\sqrt{x} + i(c_2g_L - g_0)}{\lambda -i\sqrt{x}}+\frac{\sqrt{x} - i(c_2g_L - g_0)}{\lambda +i\sqrt{x}}
\right) \cdot G \left( \frac{x + g_0^2 + g_L^2 - 2c_2g_Lg_0}{2g_Lc_1}\right)dx .
\end{eqnarray*}
Further, we put $x = \omega ^2 $ to obtain
\begin{eqnarray*}
\lambda + \frac{1}{\tau} = -\frac{\mu }{4\pi g_L c_1}\! \int^\infty_{0}\! \left( 
\frac{\omega  + i(c_2g_L - g_0)}{\lambda -i\omega }+\frac{\omega  - i(c_2g_L - g_0)}{\lambda +i\omega }
\right)  \widetilde{G}(\omega ) \omega d\omega,
\end{eqnarray*}
where $\widetilde{G}(\omega )$ is defined by Eq.(\ref{2-15}). 
Note that $\widetilde{G}(\omega )$ is an even function.
Thus, putting $\omega \mapsto -\omega $ only for the second term gives
\begin{eqnarray}
\lambda + \frac{1}{\tau} = -\frac{\mu }{4\pi g_L c_1}\! \int^\infty_{-\infty}\! 
\frac{\omega  + i(c_2g_L - g_0)}{\lambda -i\omega } \omega  \widetilde{G}(\omega ) d\omega,
\label{5-3}
\end{eqnarray}
Finally, by using the relation
\begin{eqnarray*}
\frac{\omega ^2}{\lambda -i\omega } = i\omega - \frac{i\omega \lambda }{\lambda -i\omega },
\end{eqnarray*}
we obtain
\begin{eqnarray} 
\lambda + \frac{1}{\tau} = \frac{\mu}{4\pi g_L c_1} (\lambda + g_0 - c_2 g_L)
\int^\infty_{-\infty} \frac{i\omega }{\lambda -i\omega } \widetilde{G}(\omega )d\omega .
\label{5-4}
\end{eqnarray}
When $\mu = 0$, there are no interactions of neurons and de-synchronous state should be asymptotically stable.
Thus, one may expect the existence of eigenvalues on the left half plane.
Unfortunately, this is not true.
\\

\textbf{Lemma \thedef.}
When $\mu=0$, there are no eigenvalues with nonzero real parts.
\\

\textbf{Proof.}
The right hand side of Eq.(\ref{5-4}) is a bounded holomorphic function in $\lambda $ on the right and left half planes.
Thus, a possible eigenvalue for $\mu = 0$ is only $\lambda = -1/\tau$.
However, it is not a true eigenvalue because 
it is embedded in the continuous spectrum and the corresponding eigenvector does not belong to $\mathcal{H}$.
\\

\textbf{Lemma \thedef.} Suppose $\mu > 0$.
\\
(i) If $\lambda $ is an eigenvalue, so is its complex conjugate.
\\
(ii) There are no eigenvalues on the real axis.

This lemma implies that if a bifurcation occurs at some $\mu > 0$, it will be a Hopf bifurcation.
\\

\textbf{Proof.} Putting $\lambda  = x + iy$ in (\ref{5-3}), we obtain
\begin{eqnarray*}
\left\{ \begin{array}{l}
\displaystyle x+ \frac{1}{\tau} = -\frac{\mu }{4\pi g_L c_1} \int^{\infty}_{-\infty}\! \frac{x\omega + (c_2g_L-g_0)(y-\omega )}{x^2+(y-\omega )^2}
\omega \widetilde{G}(\omega )d\omega    \\[0.4cm]
\displaystyle y = -\frac{\mu }{4\pi g_L c_1} \int^{\infty}_{-\infty}\! \frac{\omega ^2- y\omega + (c_2g_L-g_0)x}{x^2+(y-\omega )^2}
\omega \widetilde{G}(\omega )d\omega.  \\
\end{array} \right.
\end{eqnarray*}
Since $\widetilde{G}$ is an even function, it is easy to verify that if $(x,y)$ is a solution, so is $(x, -y)$.
Further, $y=0$ always satisfies the second equation.
Then, the first equation gives
\begin{eqnarray*}
x+ \frac{1}{\tau} = -\frac{\mu }{4\pi g_L c_1} \int^{\infty}_{-\infty}\! \frac{x - c_2g_L+g_0}{x^2+\omega ^2}
\omega^2 \widetilde{G}(\omega )d\omega 
\end{eqnarray*}
If $x>0$, the left hand side is positive, while the right hand side is negative.
If $x\leq 0$, it is embedded in the continuous spectrum and not a true eigenvalue. 


\section{Stability analysis 2 (generalized spectral theory)}

In Section 5, the spectrum of the operator $T$ on 
$\mathcal{H} = L^2 (\mathbf{R}, G(I)dI) \oplus L^2 (\mathbf{R}, G(I)dI) \oplus \mathbf{R}$ is conisdered:
The operator $T$ has the continuous spectrum on the imaginary axis and there are no eigenvalues on the left half plane when $\mu = 0$.
This implies that the de-synchronous state is neutrally stable but not asymptotically stable in the topology of $\mathcal{H}$.
However, the unknown function $P(t, \theta , I)$ of Eq.(\ref{2-9}) is not a $L^2$ function but a distribution.
Thus, it is better to study the dynamics in a weaker topology rather than that of $\mathcal{H}$.
For this purpose, we employ the generalized spectral theory in Section 6.
The generalized spectral theory was developed by Chiba \cite{Chi2, Chi3} to treat problems related to
continuous spectra.
He defined generalized eigenvalues and proved that it plays a similar role to usual eigenvalues.
In this section, a brief review of the generalized spectral theory is given.
All proofs are included in \cite{Chi3}.
Then, we will show that the operator $T$ has the generalized eigenvalues and they induce the stability and 
bifurcations of the de-synchronous state.

For a given function $f(\omega )$ on $\mathbf{R}$, let us consider the function 
\begin{eqnarray*}
A(\lambda ) = \int_{\mathbf{R}}\! \frac{1}{\lambda -i\omega } f(\omega )d\omega
\end{eqnarray*}
of $\lambda \in \mathbf{C}$.
It is holomorphic on the right half plane, though singular on the imaginary axis in general
because of the factor $(\lambda -i\omega )^{-1}$.
However, if $f(\omega )$ is a ``nice" function, $A(\lambda )$ may be well-defined on the imaginary axis.
The following lemma known as Sokhotskii formula~\cite{Gakhov66} is fundamental.
\\

\textbf{Lemma \thedef.} If $f(\omega )$ is a holomorphic function around the imaginary axis, $A(\lambda )$
is well-defined on the imaginary axis and it has an analytic continuation from the right half plane to the 
left half plane given by
\begin{equation}
A(\lambda ) = \left\{ \begin{array}{ll}
\displaystyle \int_{\mathbf{R}}\! \frac{1}{\lambda -i\omega } f(\omega )d\omega & (\mathrm{Re}(\lambda ) > 0) \\[0.4cm]
\displaystyle \lim_{\mathrm{Re}(\lambda ) \to 0+}\int_{\mathbf{R}}\! \frac{1}{\lambda -i\omega } f(\omega )d\omega & (\mathrm{Re}(\lambda ) = 0) \\[0.4cm]
\displaystyle \int_{\mathbf{R}}\! \frac{1}{\lambda -i\omega } f(\omega )d\omega + 2\pi f(-i\lambda ) & (\mathrm{Re}(\lambda ) < 0). \\
\end{array} \right.
\end{equation}
In \cite{Chi3}, this idea is applied not to functions but to linear operators.
Let $T$ be a linear operator defined on a Hilbert space $\mathcal{H}$ having the continuous spectrum
on the imaginary axis.
By the definition of the spectrum set, the resolvent operator $(\lambda - T)^{-1}$ is singular on the imaginary axis.
To consider the analytic continuation of $(\lambda - T)^{-1}$, let $X \subset \mathcal{H}$ be a ``nice" dense subspace of 
$\mathcal{H}$ and $X'$ be its dual space (the set of continuous linear functionals on $X$).
If a topology of $X$ is stronger than that of $\mathcal{H}$, the dual space is larger than $\mathcal{H}$ and 
we have a triplet $X\subset \mathcal{H} \subset X'$, which is called the Gelfand triplet.
We can prove for some class of operators that if the resolvent $(\lambda - T)^{-1}$ is regarded as an operator from
$X$ into $X'$, it has an analytic continuation from the right half plane to the left half plane beyond the 
continuous spectrum on the imaginary axis.
If the analytic continuation has a singularity on the left half plane, it is called a generalized eigenvalue.
Namely, if the domain of the resolvent is restricted to a ``nice" space $X$ (which depends on the problem at hand.
See \cite{Chi3, Chi4} for several examples),
then the continuous spectrum disappears. Instead of the continuous spectrum, generalized eigenvalues appear
on the left half plane.
An associated eigenvector of a generalized eigenvalue is not included in $\mathcal{H}$ but an element of the dual space $X'$.
Thus, we cannot find a generalized eigenvalue in the Hilbert space theory.
A generalized eigenvalue plays a similar role to a usual eigenvalue.
If it exists on the left half plane, it induces an exponential decay of a solution of a linear system $du/dt = Tu$ with 
respect to the topology of $X'$, which is weaker than that of $\mathcal{H}$.

For our operator $T$ given in (\ref{3-3}), the resolvent has an analytic continuation to the left half plane 
if the domain is restricted to a certain space of holomorphic functions.
The formula to give generalized eigenvalues is easily obtained from (\ref{5-4}) by applying Lemma 6.1.
For it, we assume that the function $G(I)$ defined on $\mathbf{R}$ has a continuation to the complex plane
and regard it as a function on $\mathbf{C}$.
\\

\textbf{Proposition \thedef.} 
Suppose that $G(I)$ has a meromorphic continuation to the complex plane.
The analytic continuation of the eigen-equation (\ref{5-4})
from the right half plane to the left half plane is given by 
\begin{eqnarray}
& & \lambda + \frac{1}{\tau} =  \nonumber \\
& &\!\!\!\!\!\!\!\!\! \left\{ \!\!\!\begin{array}{ll}
\displaystyle \frac{\mu}{4\pi g_L c_1} (\lambda + g_0 - c_2 g_L)
\int \frac{i\omega }{\lambda -i\omega } \widetilde{G}(\omega )d\omega & (\mathrm{Re}(\lambda) > 0)  \\[0.4cm]
\displaystyle \frac{\mu}{4\pi g_L c_1} (\lambda + g_0 - c_2 g_L) \left( 
\int \frac{i\omega }{\lambda -i\omega } \widetilde{G}(\omega )d\omega + 2\pi \lambda \widetilde{G}(-i\lambda ) \right) & (\mathrm{Re}(\lambda) < 0) \\
\end{array} \right.
\label{6-2}
\end{eqnarray}
A root of the first line is a usual eigenvalue, while that of the second line gives a generalized eigenvalue of $T$.
It is easy to confirm that if $\lambda $ is a root, then so is the complex conjugate $\overline{\lambda }$.
\\

\textbf{Theorem \thedef.} Suppose that $G(I)$ has a meromorphic continuation to the complex plane without poles on the real axis.
\\[0.2cm]
(i) When $\mu = 0$, $T$ has generalized eigenvalues $\lambda =\lambda (\mu)$ on the left half plane.
One of them is $\lambda = -1/\tau$, 
and the others are determined by poles of the function $\widetilde{G}(\omega )$ on the upper half plane.
\\
(ii) Suppose that $\eta > g_L/(2c_1)$ and $1/\tau < -c_2 g_L$.
If the variance of $G(I)$ is sufficiently small, generalized eigenvalues $\lambda (\mu)$
cross the imaginary axis transversely from left to right, and become usual eigenvalues.
\\
(iii) If $\mu$ is sufficiently large, there are no (generalized) eigenvalues on the region
$\{ \lambda \, | \, \mathrm{Re} (\lambda ) \geq -1/\tau \}$.
In particular, there are no eigenvalues on the right half plane.
\\

See Fig.\ref{fig2} (b) for a schematic picture of the motions of generalized eigenvalues.
This result implies that a pair of (generalized) eigenvalues crosses the imaginary axis at least twice,
so that a Hopf bifurcation occurs twice (see also Lemma 5.2).
At the first bifurcation, the stable synchronous firing state (gamma oscillation) bifurcates from the de-synchronous state,
and at the second bifurcation, it disappears and the de-synchronous state becomes stable again.
It is remarkable that if the edge density $\mu$ of the network is too large, synchronization does not occur.
\\

\textbf{Proof.} (i) $\lambda = -1/\tau$ is a trivial solution of (\ref{6-2}) when $\mu = 0$.
To find another solution for $\mu = 0$, note that if $\widetilde{G}(\omega )$ has a pole $\omega _*$ on the upper half plane, 
then $i\omega _*$ is a pole of $\widetilde{G}(-i\lambda )$ on the left half plane.
Let us express the Laurent series as
\begin{eqnarray*}
\widetilde{G}(-i\lambda ) = \frac{d_{p}}{(\lambda - i\omega _*)^p} + \frac{d_{p-1}}{(\lambda - i\omega _*)^{p-1}}+\cdots 
\end{eqnarray*}
for some positive integer $p$.
Substituting it into the second line of (\ref{6-2}) and multiplying the factor 
$(\lambda -i\omega _*)^p$ give
\begin{eqnarray*}
\frac{\mu}{4\pi g_L c_1} (\lambda + g_0 - c_2 g_L)\cdot 2\pi \lambda d_{p} +O(\lambda -i\omega _*) = 0,
\end{eqnarray*}
as $\lambda \to i\omega _*$.
Here note that the integral in (\ref{6-2}) is bounded at 
$\lambda =i\omega _*$ because of Lemma 6.1.
This proves that $i\omega _*$ is a solution for $\mu = 0$. 

(ii) Since $G(I-\eta)$ is an even and unimodal function around the average $\eta$, it converges to the 
delta function supported at $\eta$ as the variance tends to zero with respect to the topology of Schwartz distributions.
Thus, Eq.(\ref{5-4}) converges to the eigen-equation for the delta function (\ref{4-1}).
Fix two values $0<\mu_1 <\!< 1$ and $1 <\!<\mu_2 < \infty$.
On the closed interval $[\mu_1, \mu_2]$, the trajectory of a (generalized) eigenvalue $\lambda (\mu)$
is uniformly approximated by that of the eigenvalue for the delta function case if the variance is sufficiently small.
Then, the statement follows from Proposition 4.1.

(iii) Eq.(\ref{5-4}) is also written as
\begin{eqnarray*}
\lambda + \frac{1}{\tau} = \frac{-\mu}{4\pi g_L c_1} (\lambda + g_0 - c_2 g_L)
\int^\infty_{-\infty} \left( 1 - \frac{\lambda }{\lambda -i\omega } \right) \widetilde{G}(\omega )d\omega.
\end{eqnarray*}
By Proposition 2.2, $g_0(\mu) \to \infty$ as $\mu \to \infty$.
By changing the variable as $\omega = g_0 \hat{\omega }$, the above integral is rewritten as
\begin{eqnarray*}
g_0 \int^\infty_{-\infty} \left( 1-\frac{\lambda }{\lambda -ig_0 \hat{\omega }} \right)
G\left(\frac{g_0^2(\hat{\omega }^2 + 1)+g_L^2-2c_2g_Lg_0}{2g_Lc_1} \right) d\hat{\omega }.
\end{eqnarray*}
This shows that the second term $\int^\infty_{-\infty} \lambda /(\lambda -i\omega ) \widetilde{G}(\omega )d\omega$
is infinitesimally smaller than the first term $\int^\infty_{-\infty} \widetilde{G}(\omega )d\omega$
as $g_0 \to \infty$.
Hence, considering only leading terms as $\mu \to \infty$, we obtain
\begin{eqnarray*}
\lambda + \frac{1}{\tau} = \frac{-\mu}{4\pi g_L c_1} (\lambda + g_0)
\int^\infty_{-\infty} \widetilde{G}(\omega )d\omega + o(\mu).
\end{eqnarray*}
If $\mathrm{Re} (\lambda ) > -1/\tau$, 
the real part of the left hand side is positive, while that of the right hand side is negative. 
This proves that there are no eigenvalues on the region $\mathrm{Re} (\lambda ) > -1/\tau$.
The same argument is also valid for the second line of (\ref{6-2}).
Thus, there are no generalized eigenvalues on the same region.
\\

\textbf{Remark.} If $\widetilde{G}(\omega )$ does not have poles on the upper half plane,
there are no generalized eigenvalues except $-1/\tau$ when $\mu = 0$.
Then, the statement (ii) implies that generalized eigenvalues pop out from the infinity $\infty$ to the left half plane when $\mu = +0$,
and they approach to the imaginary axis as the variance tends to zero for fixed $\mu > 0$.

Although the Lorentzian distribution $G(I) = (\Delta /\pi )/ ((I-\eta)^2 + \Delta^2) $ does not have the variance, the statement (ii) is applicable 
because it converges to the delta function as $\Delta \to 0$.

\section{Numerical results}
\begin{figure}
\begin{center}
\includegraphics[scale=0.8]{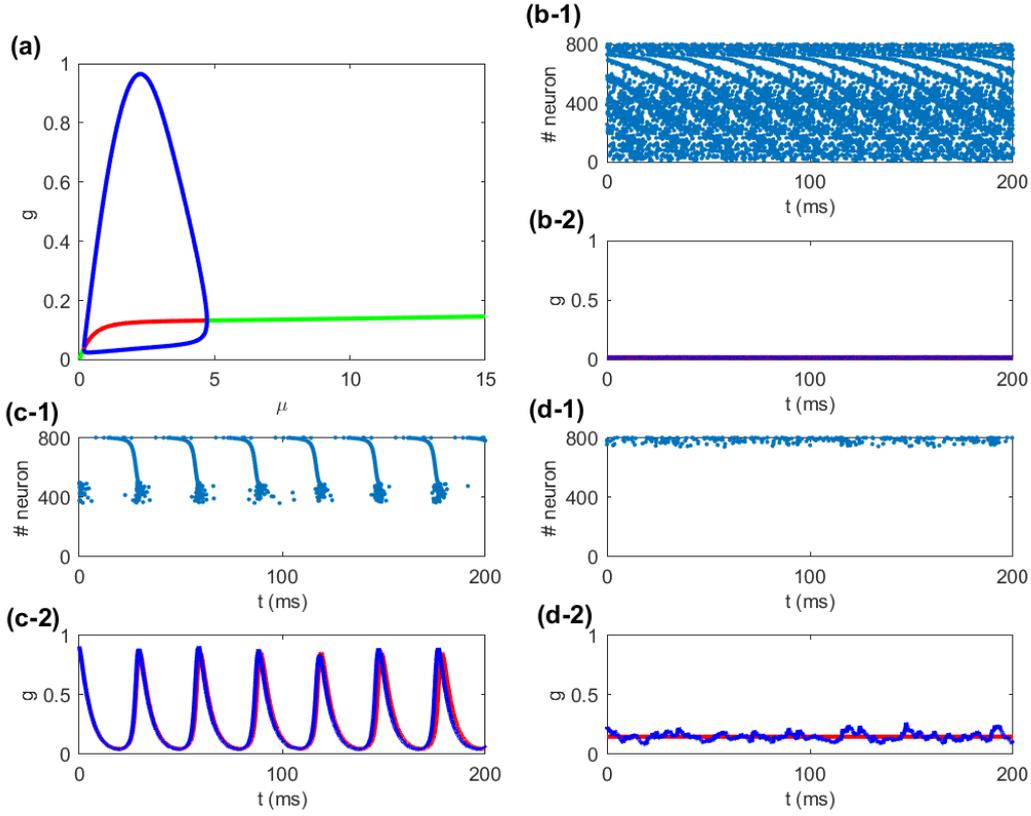}
\caption[]{(Color online)(a) Bifurcation diagram of Eq.~(\ref{7-2}). The green line is a stable steady state and the red one is an unstable one. The blue lines indicate the upper
and lower bounds of the gamma oscillation.  (b-1) Raster plot with $\mu=8.6\cdot 10^{-2}$, 
(b-2) Time-course of the conductance $g$ with $\mu=8.6\cdot 10^{-2}$. 
The blue line is a result of the individual neuron model (Eq.~(\ref{2-5}) and (\ref{2-7})) and the red line is that of the macroscopic equation  (Eq.~(\ref{7-2})). 
(c-1) Raster plot with $\mu=3.2$, (c-2) Time-course of $g$ obtained by the neuron model (blue) and the macroscopic equation (red) with $\mu=3.2$.  
 (d-1) Raster plot with $\mu=15.0$, (c-2) Time-course of $g$ obtained by the neuron model (blue) and the macroscopic equation (red) with $\mu=15.0$.  
}
\label{fig3}
\end{center}
\end{figure}

Here we consider the Lorentzian distribution $G(I) = (\Delta /\pi )/ ((I-\eta)^2 + \Delta^2) $.
Then, the reduced equation (Eq.~(\ref{3-2})) can be further simplified by virtue of the residue theorem.  
We also introduce the order parameter $\alpha(t) \in \mathbf{C}$ as 
\begin{equation}
\alpha (t)=\int_{\mathbf{R}} \int_{0}^{2\pi} P(t,\theta, I) e^{i\theta} G(I) d\theta dI.
\end{equation}
Ott Antonsen ansatz and the residue theorem yield the following relations 
\begin{eqnarray} 
\alpha(t)&=&z_1(t, \eta+i\Delta), \\ 
A(t) &=& \frac{g_L}{2\pi} \left[ 1+2 {\rm Re}\left(\frac{-\alpha(t)}{1+\alpha(t)} \right)\right].
\label{7-1}
\end{eqnarray}
By using these relations, the dynamics of the order parameter and the synaptic conductance are described as 
\begin{eqnarray} 
\left\{ 
\begin{array}{l}
\dot{\alpha}= i(\, f(\eta+i\Delta) \alpha^2 + h(\eta+i\Delta) \alpha + \overline{f(\eta+i\Delta)}\,)  \\[0.2cm]
\displaystyle \dot{g}_{syn} = -\frac{1}{\tau}g_{syn} + \mu A(t) \\[0.2cm]
\displaystyle A(t) = \frac{g_L}{2\pi} \left[ 1+2 {\rm Re}\left(\frac{-\alpha(t)}{1+\alpha(t)} \right)\right],
\end{array} \right.
\label{7-2}
\end{eqnarray}
which has been presented in Ref.~\cite{Aki}. 
We employ bifurcation analyses for the equations. Here, we adopt $(\eta, \Delta)=(2, 0.05)$ and set $\mu$ as a bifurcation parameter. 
The bifurcation diagram is shown in Fig.~\ref{fig3}(a). 
We can see that as $\mu$ increases, the populational gamma oscillation emerges at $\mu \sim 0.18$ via Hopf bifurcation. 
The oscillatory state disappears by further increase of $\mu$ at $\mu \sim 4.7$. 

These dynamics are confirmed by direct numerical computations of the individual neuron (Eq.~(\ref{2-5})) with homogenized synaptic dynamics (Eq.~(\ref{2-7})) to avoid the effects of random variation and finite size effect that appear the individual network structure.
We show raster plots of individual firings with $\mu=8.6\cdot 10^{-2}, 3.2$, and 15.0 in Figs.~\ref{fig3}(b-1),(c-1), and (d-1), respectively. 
The vertical axis is a neuron index that is sorted in ascending order of $I$.
We can see that the emergence and disappearance of the populational oscillation coincides with the bifurcation analysis of the macroscopic equation (Eq.~(\ref{7-2})). 
Additionally, we show time-course of $g$ with $\mu=8.6\cdot 10^{-2}, 3.2$, and 15.0 in Figs.~\ref{fig3}(b-2),(c-2), and (d-2), respectively. 
The ferquency of the oscillation in Figs.~\ref{fig3}(c-2) is 34~Hz, therefore,  it is gamma frequency oscillation. 
Numerical results for the individual neuron model (Eq.~(\ref{2-5}) and (\ref{2-7})) well match to those for the macroscopic equation (Eq.~(\ref{7-2})).

\section{Discussion}
For the analysis of populational neuronal firings with steady-states or slow frequency oscillations, 
it is effective to use the firing rate model, such as Wilson-Cowan model~\cite{WC}, derived by the adiabatic approximation. 
On the other hand, for analysis of oscillations with high frequency such as gamma oscillations, 
the firing rate model is not appropriate as an approximation~\cite{Ermentrout2010}.
Instead, Vlasov equation of the MT model posesses the dynamics of distribution itself. 
This is why we utilized the MT model for the analyses of gamma oscillations. 
Although some studies have analyzed the coupled dynamics of the conventional theta neuron model~\cite{Paul1, Paul2, Gutkin, Kilpatrick}, 
the strength of using the MT model lies in its ability to evaluate the collective dynamics under synaptic interactions with physiologically appropriate conductance and reversal potential.

The inhibitory synaptic interactions by GABAergic neurons are known to generate a macroscopic oscillation in the gamma frequency range, which is named ING (interneuron gamma oscillation)~\cite{Wang0, Bartos}. 
In addition, synaptic connectivity in the cerebral cortex is different from individual to individual and depends on the developmental stage.
It has been reported that the synaptic connections increase in early phase of development followed by decrease 
by activity-dependent synaptic pruning~\cite{Huttenlocher, Uesaka, Wu}. 

Therefore, we investigate the relationship between the synaptic connectivitiy and macroscopic gamma oscillations 
in the population of the inhibitory neurons under meromorphic function $G(I)$ that is even and unimodal around the average and has no poles on the real axis. 
If the variance of $G(I)$ is sufficiently small, the generalized spectral theory indicate that 
the populational gamma oscillation emerges only when the synaptic connectivity has a value within an approariate range, 
which is confirmed by the numerical bifurcation analysis under the Lorentzian distribution for $G(I)$. 
As shown in Fig.~\ref{fig3}(c-2), the macroscopic oscillation with frequency of 33.6~Hz is observed, which is within the range of gamma oscillation (30--80 Hz). 
In addition, the dynamics of two asynchronous states by the numerical computations (Fig.~\ref{fig3}(b-1) and (d-1)) look very different.
In the asynchronous state with smaller $\mu$, almost every neuron fires aperiodically (Fig.~\ref{fig3}(b-1)). 
On the contrary, in the asynchronous state with larger $\mu$, only partial population fires and others keep silent (Fig.~\ref{fig3}(b-1)).
We could say that normal neuronal populations avoid the latter state that has excess connections between inhibitory neurons. 
Besides biological plausibility, the second bifurcation seems mathematically meaningful. The generalized spectral theory indicates
that these transitions of gamma oscillations are introduced by the nontrivial motion of the pair of the generalized eigenvalues that cross the imaginary axis twice. 
Remarkably, both types of asynchronous state are well characterized by the generalized eigenvalues on the left half plane. 

\section*{Acknowledgment}
This study is supported in part by JST PRESTO (JPMJPR14E2) to KK, and by JST PRESTO (JPMJPR16E7) to HC.


\end{document}